\documentclass[article]{aa}
\usepackage{graphicx}
\usepackage{natbib}
\bibpunct{(}{)}{;}{a}{}{,}

\begin{document}

\hyphenation{Fe-bru-ary Gra-na-da mo-le-cu-le mo-le-cu-les}
     \title{Estimation and reduction of the uncertainties in chemical models: Application to hot core chemistry}

     \subtitle{}

     \author{V. Wakelam\inst{1}, F. Selsis\inst{2}, E. Herbst\inst{1,3}, P. Caselli\inst{4}}
     \institute{ Department of Physics, The Ohio State University, Columbus, OH 43210, USA \and Centre de Recherche Astronomique de Lyon, Ecole Normale Sup\'erieure, 46 All\'ee d'Italie, F-69364 Lyon cedex 7, France \and Departments of Astronomy and Chemistry, The Ohio State University, Columbus, OH 43210, USA \and INAF-Osservatorio Astrofisico di Arcetri, Largo E. Fermi, 50125 Firenze, Italy      }
     \offprints{wakelam@mps.ohio-state.edu}
     \date{Received 21 June 2005 / Accepted 25 August 2005 }
     
     \abstract{ It is not common to consider the role of uncertainties in the rate coefficients used in interstellar gas-phase chemical models.  In this paper, we report a new method to determine both the uncertainties in calculated molecular abundances and their sensitivities to underlying uncertainties in the kinetic data utilized.  The method is used in hot core models to determine if previous analyses of the age and the applicable cosmic-ray ionization rate are valid. We conclude that for young hot cores ($\le 10^4$~yr), the modeling uncertainties related to rate coefficients are reasonable so that comparisons with observations make sense. On the contrary, the modeling of older hot cores is characterized by strong uncertainties for some of the important species. In both cases, it is crucial to take into account these uncertainties to draw conclusions from the comparison of observations with chemical models. 
    }

     \titlerunning{Chemical modeling uncertainties}
     \authorrunning{Wakelam et al.}

     \maketitle

\section{Introduction}

Chemical models usually include thousands of reactions. The rate coefficients of these
reactions are known to have uncertainties that affect the theoretical abundances computed by 
models. Estimating the resulting error bars is the only rigorous way to compare modeled and 
observed abundances. Although such an estimation of uncertainties has been undertaken in other fields where chemical networks are used, e.g. in atmospheric photochemistry \citep[see][]{2003A&A...398..335D}, this important aspect has been relatively neglected in interstellar astrochemistry.  To the best of our knowledge, only three previous studies have appeared.  In 1976, \citeauthor{1976ApJ...209..116K} published a sensitivity analysis of  molecular abundances to  uncertainties in rate coefficients under steady-state conditions in  dark clouds. The main purpose of this study was more the identification of chemical schemes forming the species than an estimate of "real" uncertainties in abundance.  The two more recent works are by
(i) \citet{Roueff1996}, who studied uncertainties in chemical modeling in the region of bistability,
%, while \citet{Markwick2002} reported a study for cold dark clouds in an oral presentation. 
and (ii) \citet{2004AstL...30..566V}, who also performed a study for the steady-state chemistry occurring in  cold  dark clouds. 
%These authors used a method in which the error in the rate coefficients is contained in an equi-probable distribution and applied it at steady state. 
Independently we developed some procedures to include in a systematic manner the imprecision due to chemical data uncertainties in our interstellar chemical models.  
%In our approach, the uncertainties in rate coefficients possess standard normal (Gaussian) distributions and 
In addition, a sensitivity analysis is utilized to focus on the major reactions that determine the uncertainties. 
In this paper,  we study the implications on the modeling and dating of protostellar hot cores.

%However, their method overestimates the uncertainty of the reaction rates since they
%use an equiprobable distribution of the error whereas a gaussian one is more correct (ref?). 

During the collapse of a protostar, the temperature and density nearest the center increase  to form a 
hot-core region where the dust mantles evaporate, releasing molecules into the gas phase. Considering the time of evaporation of the mantles as the initial time, hot cores can be dated by their  
 chemical evolution.  
Indeed, the abundances of some chemical compounds evolve more rapidly than the structure of the protostellar envelope. Therefore, the comparison of the time-dependent  abundances computed for these species with the observed ones can constrain the age of the
source. Sulphur-bearing species are usually considered to be good chemical clocks for hot cores: 
sulphur chemistry is initiated by evaporation from icy mantles and evolves sufficiently rapidly for the purpose \citep{1997ApJ...481..396C,1998A&A...338..713H,2004A&A...422..159W}.
The chemical network and model we use for this study has been previously applied to sulphur chemistry in the hot corino\footnote{\emph{Hot corino} refers to the hot core of a low mass protostar \citep{2004ApJ...617L..69B}. } of IRAS16293-2422 in order to
constrain both the age of the source and the form of sulphur
evaporated from the grains \citep{2004A&A...422..159W}. The observed
abundance ratios of sulphur-bearing species (SO$_2$/SO, SO$_2$/H$_2$S and OCS/H$_2$S) were reproduced, to within the observational error bars, only for an age of $1500-2500$~yr, with the added assumption  that the main source of evaporated sulphur is in the atomic 
form. We have now revisited these results by including uncertainties in the chemical rate coefficients. 

In this article, we first describe the general method to estimate the uncertainties in computed abundances and to identify the reactions of the chemical network that are mainly responsible for these uncertainties. Then, we show how the imprecision of the model results affects the scientific conclusions for the specific case of  hot core chemistry. In particular, we explore the consequences in the use of molecular abundances to constrain the cosmic-ray-induced ionization rate and the age of a hot core.

\section{Chemical modeling}
\subsection{Model Description}
We have used the pseudo-time-dependent chemical model described in
\citet{2004A&A...422..159W}. This code computes the chemical evolution
of gas-phase species at a fixed gas temperature and density and for
initial molecular abundances.  The model
includes 930 reactions involving 77 species that contain the elements H, He, C, O and S. The most complex molecule in our reduced sample is protonated methanol (CH$_3$OH$_{2}^{+}$), which contains seven atoms. The model includes 780 gas-phase reactions (ion-neutral, neutral-neutral and dissociative recombination processes), and
allows species to stick to the surfaces of dust particles, as well as to
evaporate  and undergo charge exchange with the grains. The chemical network originally used for the model was taken from the osu.2003 network\footnote{http://www.physics.ohio-state.edu/$\sim$eric/research\_files/ cddata.Jul04.op} \citep{2004MNRAS.350..323S}. 
For this study, however, we mainly used the rate coefficients given
in the UMIST99 database\footnote{http://www.rate99.co.uk} because it provides estimates for uncertainties in the reaction rate coefficients (see next section). This is the reason for small differences between the calculated abundances given in \citet{2004A&A...422..159W} and in the present study. We anticipate that these slight discrepancies remain within the estimated error at the considered temperatures. 

Following \citet{2004A&A...422..159W}, the initial gas phase composition is computed from the model for a molecular cloud with a temperature of 10~K and an H$_2$ density of $10^4$~cm$^{-3}$. The initial time ($t$=0) corresponds to the formation of the hot core, when the temperature increases to 100~K (the evaporation temperature of H$_2$O), injecting into the gas phase the grain-mantle composition. We took the mantle composition observed in the environment of massive protostars \citep[see][]{2004A&A...422..159W}. 
The abundance and form of sulphur initially evaporated from the grains are not quite known and  remain a serious problem \citep[see][]{1999MNRAS.306..691R,2004A&A...422..159W}. However,  the species OCS and H$_2$S  seem to be present in the grain mantles. Indeed the first one has been observed in the solid state with a low fractional abundance of $10^{-7}$ compared with H$_2$  \citep{1997ApJ...479..839P}. Although the second one has never been detected on grains, the large abundances ($>10^{-8}$ with respect to H$_2$)  observed in hot cores and along outflows cannot be produced by gas phase routes (which accounts for less than 1\% of the observed abundances). Thus, we assume that H$_2$S is present on grains with an abundance lower than or equal to $10^{-7}$ with respect to H$_2$ \citep{1998ARA&A..36..317V}.  In addition to these molecules, \citet{2004A&A...422..159W} showed that the only possibility to reproduce the sulphur-bearing abundances observed in the low mass hot core of IRAS16293-2422 is to evaporate a large amount of atomic sulphur. Based on this prior analysis, we used a mantle composition here in which  H$_2$S, OCS and S have abundances (with respect to H$_2$) of $10^{-7}$, $10^{-7}$, and $3\times 10^{-5}$, respectively. 

\subsection{Method}\label{method}
\begin{figure}
\centering
\includegraphics[angle=90,width=0.7\linewidth]{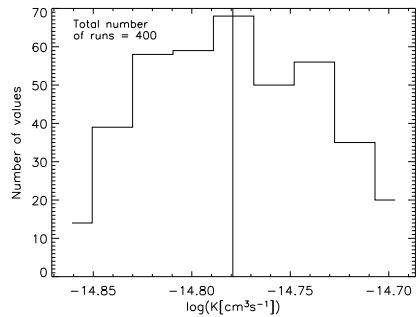}
\caption{Distribution of the rate coefficient of the reaction O + SO $\rightarrow$ SO$_2$ +
h$\nu$ for 400 runs at a temperature of 100~K. The solid vertical line is the standard rate of UMIST.}
\label{K_run}
\end{figure}

To include uncertainties in kinetic data  and to estimate their impact on the modeled abundances, we applied a Monte-Carlo procedure developed by Dobrijevic et al. (\citeyear{1998P&SS...46..491D,2003A&A...398..335D}) to study the gas-phase chemistry of the atmospheres of giant planets. This method was itself inspired by earlier studies dedicated to terrestrial stratospheric chemistry \citep{1978JGR....83.3074S,1991JGR....9613089T,1996JGR...10120953S}.
First of all, we assume that the errors in the model results are due only to uncertainties in the gas-phase reaction rate coefficients, which implies that temperature, gas and grain densities, ionizing radiation, elemental abundances, and initial concentrations are all well characterized. This is obviously an ideal case but it allows us to evaluate the  error specifically due to the kinetic data and to indentify the main reactions generating this error. 

Each reaction $i$ included in the model possesses a rate coefficient $K_i$. The UMIST kinetic database gives us the parameters to calculate the standard set of coefficients  $K_{0,i}$ at a given temperature for bimolecular reactions, direct cosmic-ray ionization, and cosmic-ray-induced photoreactions.  The UMIST database also provides a factor $F_i$ quantifying the uncertainty of the  $K_i$. Under the physical conditions of the interstellar medium, this factor $F_i$ itself cannot be estimated to a high level of precision, and  rate coefficients in the UMIST database are classified in only 4 categories of precision:  within 25\% ($F=1.25$), 50\% ($F=1.50$), a factor of 2 ($F=2.0$) and a factor of 10 ($F=10$).  For the few reactions for which the coefficient $F$ is not available, we assume an error within 25\%. This choice is arbitrary but motivated by several arguments. To begin with, there is no obvious reason to assume that the reactions from osu.2003 but absent from UMIST are affected, on average, by a larger uncertainty than the others. Therefore, we chose the precision that dominates the set of UMIST reactions (40\% of them have $F=1.25$).  Another reason has to do with the next goal of our study, which is to identify the main reactions responsible for the error in the modeling. By assuming a high uncertainty on the reactions without known $F$ factor, our selection method would tend to point to these reactions. We prefer to avoid such bias, as we already know that these rate coefficients need to be considered with priority in order to have their uncertainty constrained. \\

We assume (as illustrated in Fig.~\ref{K_run}) that  a rate coefficient $K$ can be considered as a random variable, lognormally distributed over an uncertainty range centred on the recommended value. This basic assumption is well verified in the absence of systematic errors \citep{1991JGR....9613089T,1996JGR...10120953S}. For this reason,  kinetic databases commonly provide the uncertainty as a $\Delta \log K$ error (see for instance the IUPAC database: Atkinson et al. \citeyear{2004AtkinsonACP}) or as the factor $F$ (which is equivalent, as $\log F = \Delta \log K$). Here, our method differs from \citet{2004AstL...30..566V} since these authors used an equiprobable distribution within the error interval given in UMIST. However, the authors noticed that their results are not significantly affected by the form of the distribution. \\
Thus, given a factor $F_i$ estimated at a 1$\sigma$ level of confidence, our approach implies that the value of $\log (K_i)$ follows a normal distribution with a standard deviation $\log (F_i)$. In other words, the probability to find $K_i$ between $K_{0,i}/F_i$ and $K_{0,i} \times F_i$ is 68.3\%. 

With the standard set of rate coefficients $K_{0,i}$, we can compute the standard evolution of the abundances $X_{0,j}(t)$ of the species $j$ in the considered medium, which is what typical chemical models do. The specificity of our method consists of generating $N$ new sets of rate coefficients (typically $N=400$) by taking into account the uncertainties affecting each $K_i$. This is done by generating $N$ sets of random numbers $\epsilon_{i}$ with a normal distribution centered on 0 and with a standard deviation of 1. For each set of $\epsilon_{i}$, the new $K_i$ are given by
\begin{equation}
\log K_i = \log K_{0,i} + \epsilon_{i} \log F_i
\end{equation}
Running the model $N$ times, once for each set of rate coefficients, produces $N$ values of the abundances $X_{j}(t)$ of the species $j$ at  time $t$. The mean value of $\log X_{j}(t)$ ($\overline{\log X_j}(t)$) gives us the "recommended" value while the dispersion of $\log X_{j}(t)$ around  $\overline{\log X_j}(t)$ determines the error due to kinetic data uncertainties. \\

Once the error in the computed abundances is estimated, it is extremely useful to identify which reactions among the many possible are responsible for most of this error.  It is indeed  of high interest to be able to point out a few reactions for which more accurate measurements, or theoretical estimates, of the rate coefficient would significantly reduce the error in model results.

Let us consider a species $j$ at a time $t$. The error in its abundance due to uncertainties in kinetic data  is $\Delta \log X_{j}(t)$. Each of the $N_R$ reactions included in the model contributes to this error. In order to estimate their individual contribution to the errors, we perform $N_R$ new runs of the model. In each run $i$, we replace the standard rate $K_{0,i}$ of one reaction by its $1\sigma$ upper value $K_{0,i} \times F_i$, all the other rates being fixed to their standard values $K_0$. Each run $i$ performed with such a perturbation of the rate $K_i$ produces an abundance $X_j^i(t)$ for every molecular species $j$. The ratio $R_j^i(t)$, defined by the relation
\begin{equation}
 R_j^i(t) = \frac{|X_j^i(t)-X_{0,j}(t)|}{X_{0,j}(t)},
 \end{equation}
yields an index quantifying the influence of the reaction $i$ on the total error for species $j$. The ratio $R_j^i(t)$ can also be computed for abundance ratios instead of abundances when necessary. 

The rigorous way to estimate the individual error induced by a reaction $i$ would be to compare the error $\Delta \log X_{j}(t)$ obtained when all reaction rates are randomly chosen within their uncertainty range, with the one obtained when all the reactions but $i$ are randomly chosen. The difference betwen these two values of  $\Delta \log X_{j}(t)$ would give the error in $X_j(t)$ that is due to reaction $i$ only. However, such an approach requires $N_R \times N$ runs ($780 \times 400 = 312,000$ in our case), which is not affordable in terms of computational time and data storage. Using the index $R_j^i(t)$ is an alternative practical method although, in some cases, it may not be sufficient to identify some of the reactions having a significant impact on the error. This is due to the non-linearity of chemical networks and the complex structure of the routes to form some species. Nevertheless, once we have selected a group of reactions with the highest index $R_j^i(t)$ we can artificially nullify or reduce their uncertainty and check the resulting  decrease of the error $\Delta \log X_{j}(t)$. It is important to note that a set of reactions identified to be responsible for most of the error in the abundance of one given compound $j$ at a time $t$ can have a minor effect on the error in the abundance of another species and/or at another time.

 Our approach can be compared with the earlier work of  \citet{1976ApJ...209..116K}, who studied the sensitivity of molecular abundances to  reactions involved in dark-cloud chemistry. Their chemical network consisted of 19 species and 31 reactions, allowing them to use a Fourier analysis technique \citep[FAST,][]{Cukier1973},  which consists of periodically varying all the rate coefficients and solving the linear system as a function of the coefficients. This method is efficient for small systems but the number of required solutions increases steeply with the number of reactions. With our network of 780 reactions, this technique would require much more computation time than the Monte Carlo method we applied, without providing more information.  Moreover, FAST is applicable only at steady state, which is not reached in astronomical objects such as dark clouds and hot cores, where time dependence has to be included. 

\section{Results of Uncertainty Calculations}

In this section, we present general results concerning the uncertainties of computed abundances (with respect to H$_2$) in hot cores for typical parameters: a temperature of 100~K, a density of $10^7$~cm$^{-3}$, and ages lower than $10^6$~yr. The two last parts of this section present the consequences of the rate coefficient uncertainties on two practical applications: the determination of the H$_2$ cosmic-ray ionization rate and the determination of the hot core age using observed chemical abundances. 

\subsection{Definition of the error}\label{def_error}

\begin{figure}
\centering
\includegraphics[width=1\linewidth]{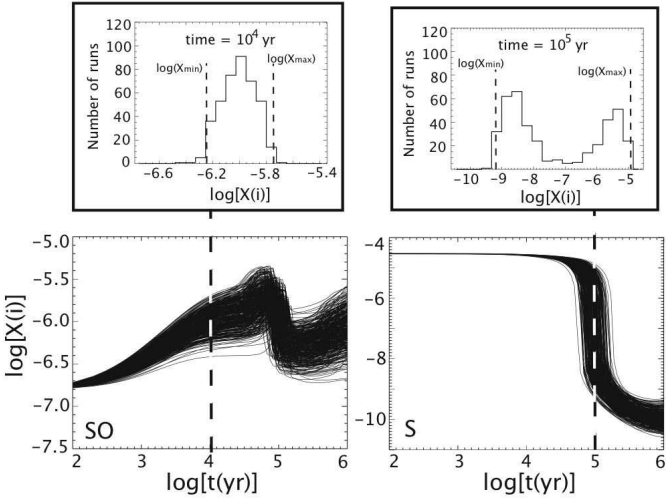}
\caption{Abundances of SO and S computed with 400 runs (lower panel) as a function of time. The upper panels represent histograms of the abundance distribution of these species for a given time ($10^4$ and $10^5$~yr for SO and S respectively).  The gas temperature is 100~K and the H$_2$ density $10^7$~cm$^{-3}$.}
\label{allrun_dis}
\end{figure}

The distribution of $\log X$ (the logarithm of the abundance with respect to H$_2$ at an integration time $t$) is well fitted, at most integration times, by a Gaussian function centred on the mean value $\overline{\log X}$ (see the abundance of SO in Fig.~\ref{allrun_dis}). In these cases, we can use the standard deviation $\sigma$ of the  $\log X$ distribution to evaluate the error $\Delta \log X$.
The probability of finding $\log X$ within $\overline{\log X} \pm \sigma$ and $\overline{\log X} \pm 2\sigma$ is 68.3\% and 95.4\%, respectively. 
%In our study, we use the 2$\sigma$ error (95.4\% confidence level). 

Close to a strong inflection point in the evolution of the abundance (i.e., a steep edge in the time derivative of the abundance), the distribution of $\log X$ is no longer Gaussian.  Indeed, modifying the rate coefficients here produces a temporal shift in the evolution of $\log X$ that can result in two separate distributions at the same time $t$, one on either side of the mean value $\overline{\log X}$.  As an example, consider the inflection in the evolution of the atomic sulphur abundance in Fig.~\ref{allrun_dis} and the resulting bimodal distribution at 10$^5$~yr. In such a case, the standard deviation seriously overestimates the error.  Rigorously, when this occurs, we should fit the distribution by two Gaussians and define two disconnected error bars but, instead, we used the following method to estimate $\Delta \log X$ at any time $t$. First, we calculate the normalized density of abundance-vs-time curves (or density of probablity) $\frac{1}{N} \frac{\delta n}{\delta \log X}$, where $\delta n$ is the number of curves per $\delta \log X$ interval and   $N$ is the total number of runs. Then we identify the smallest interval $[\log X_{\mathrm{min}},\log X_{\mathrm{max}}]$ (see Fig.~\ref{allrun_dis}) that contains 95.4\% of the curves, and we define the error as follows:
\begin{equation}\label{relat_error}
 \Delta \log X = \frac{1}{2}( \log X_{\mathrm{max}} - \log X_{\mathrm{min}})
\end{equation}
When $\log X$ does follow a Gaussian distribution, this expression reduces to $ \Delta \log X= 2\sigma$. 
{\bf In the rest of the paper, we refer to the error computed with this method, which has a level of confidence of 95.4\%.} Note again that abundance ratios can be treated the same way as abundances with respect to H$_2$.

If $\Delta \log X = \log s$, the error domain $[X_{\mathrm{min}},X{\mathrm{max}}]$ is $[\frac{X}{s},sX]$. An alternative means of interpreting the error is $[X_{\mathrm{min}},X{\mathrm{max}}] = \bar{X}_{-\Delta X_{-}}^{+\Delta X_{+}}$ with $\frac{\Delta X_{-}}{X} = 1-\frac{1}{s}$ and $\frac{\Delta X_{-}}{X} = s-1$.   When $\Delta \log X \ll 1$, $\Delta X_{-} \approx \Delta X_{+}$ and the error $\Delta \log X$ is proportional to the relative error in $X$:   $\Delta \log X \approx \frac{\Delta X}{\ln{}10 X}$. When $X$ is lognormally distributed $s=10^{2\sigma }$. As an example, if $\log X$ is known to within $\Delta \log X \pm 0.01$, the error $[\Delta X_{-}, \Delta X_{+}]$ in $X$ is about $\pm 2.3$\% of X and $[X_{\mathrm{min}},X{\mathrm{max}}]=[\frac{X}{1.023}, 1.023X]$.  Table~\ref{conversion} gives the relative error and the error domain in $X$ corresponding to some values of $\Delta \log X$ found in our simulations.

\begin{table}
\begin{center}
\caption{Conversion between $\Delta \log X$, relative abundances and error domains in X.\label{conversion}}
\begin{tabular}{lllll}
\hline
 $\Delta \log X$ & $\frac{\Delta X_{-}}{\bar{X}}$ & $\frac{\Delta X_{+}}{\bar{X}}$  & $\frac{X_{\mathrm{min}}}{X}$ &  $\frac{X_{\mathrm{max}}}{X}$\\
 \hline
      0.01 &      0.023 &      0.023 &      0.977 &      1.023 \\
      0.10 &       0.21 &       0.26 &       0.79 &       1.26 \\
      0.50 &       0.68 &       2.16 &       0.32 &       3.16 \\
      1.00 &       0.90 &       9.0 &       0.10 &      10.0 \\
      1.50 &       0.97 &      30.6 &       0.03 &      31.6 \\
      2.00 &       0.99 &      99 &       0.01 &     100 \\
  \hline
\end{tabular}
\end{center}
\end{table}

\subsection{Uncertainties in the abundances}

\begin{figure}
%\centering
\includegraphics[width=1\linewidth]{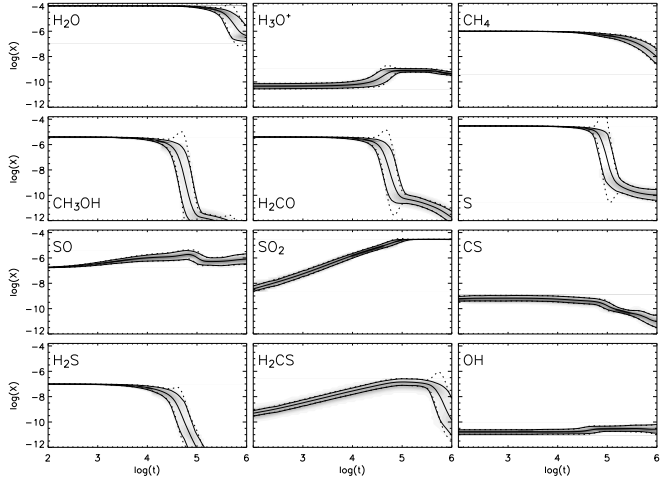}
\caption{Evolution of the fractional abundances (/H$_2$) of some species. The central solid line shows the evolution of the mean value $\overline{\log X}$ while the 2 other lines delimit the domain containing 95.4\% of the computed values of $\log X$. The dashed lines give the 2$\sigma$ standard deviation from the mean value with a Gaussian fit. Grey levels represent the density of probability (see text). }
\label{ab_deviation}
\end{figure}

\begin{figure}
\centering
\includegraphics[width=1\linewidth]{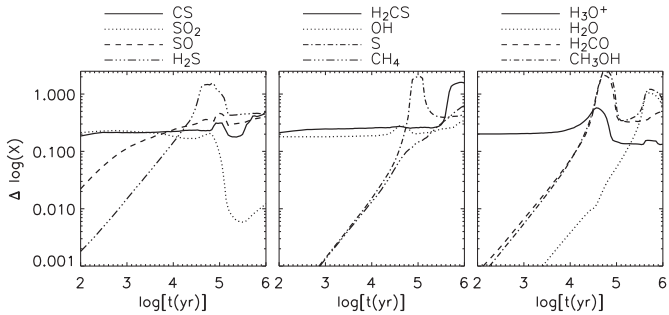}
\caption{Error $\Delta \log X$ of the abundance as a function of time for some species abundant in hot cores. }
\label{sigma_time}
\end{figure}

\begin{figure}
\begin{center}
\includegraphics[width=0.9\linewidth]{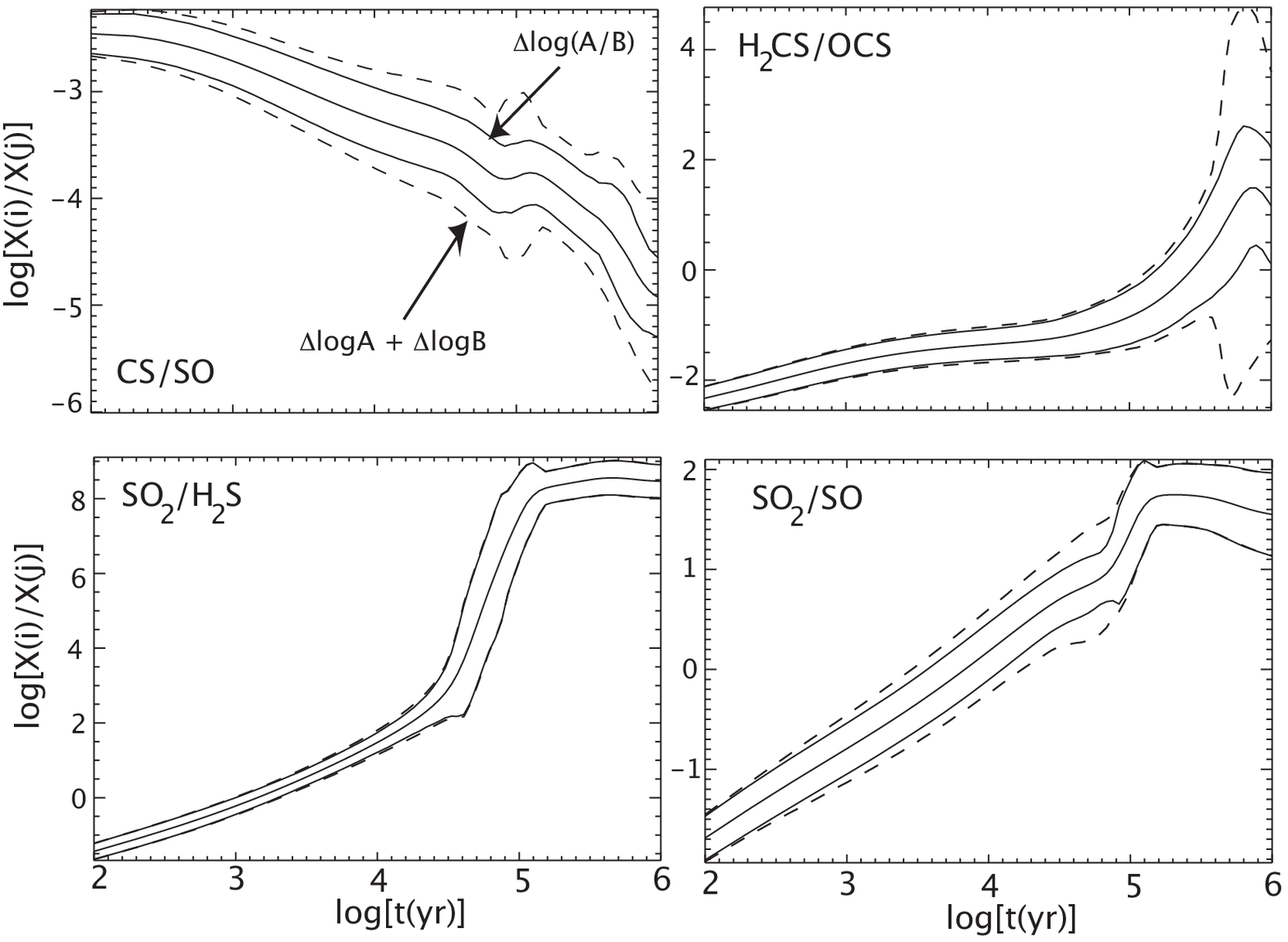}
\caption{The left panel shows the mean abundance of the SO$_2$ and 
H$_2$S molecules as a function of time (solid lines) with their error bars (grey contours) and the standard abundance (dashed grey lines). The right panel represents the ratio between the mean and the standard abundances ($\overline{X}/X_0$) for the two molecules as a function of time. }
\label{ab_mean_ref}
\end{center}
\end{figure}

 Figure~\ref{ab_deviation} shows the evolution of $\log X$ computed for a variety of species at a temperature of 100~K and a H$_2$ density  of $10^7$~cm$^{-3}$. The evolution of $\overline{\log X} \pm \Delta \log X$ (envelopes containing 95.4\% of the values as described above, solid lines on Fig.~\ref{ab_deviation}) and $\overline{\log X} \pm 2\sigma$ (standard deviation, dashed lines on Fig.~\ref{ab_deviation}) are compared. Discrepancies between  solid and dashed lines indicate a non-Gaussian distribution of $\log X$. 
 Figure~\ref{sigma_time} shows the evolution of the error $\Delta \log X$ (estimated using the method explained above, equation~\ref{relat_error}) as a function of time for some abundant species ($X\ge 10^{-11}$) under the same physical conditions.  
 
The maximum $\Delta \log X$ found in all our calculations is about 2 (for the species H$_2$CO, CH$_3$OH and S around $10^5$~yr). 
%As explained in \S~\ref{def_error}, when $\Delta \log X \gg 0.1$, the relative error in $X$ is no longer proportional to $\Delta \log X $, and the $\Delta X_{-}$ and $\Delta X_{+}$ intervals become significantly different. 
With $\Delta \log X = 2$, the abundances $X$ go from $X/100$ to $100 \times X$,  which means that the total error spans 4 orders of magnitude. However, such large error bars occur at late times when the hot core has probably ceased to exist. For times lower than $10^4$~yr, the relative error in X remains below 50\% ($\Delta \log X \le 0.2$) and is comparable to the uncertainties on observed abundances produced solely by telescope and atmospheric calibrations. 
%At these young ages ($t<10^4$~yr), one can use the approximation $\frac{\Delta X_{+}}{X}  \approx \frac{\Delta X_{-}}{X} \approx 2.3 \Delta \log X \approx 2.3 \times 2 \sigma(\log X)$ to estimate the relative error. 

The error $\Delta \log X$ in the computed fractional abundance depends on the species, as well as the integration time (see Fig.~\ref{sigma_time}), the temperature, and the density. However, general patterns can be noticed. The error is logically related to the generational rank of the molecule: the abundances of the first generation of molecules at small integration times are determined by the initial conditions and are not affected seriously by kinetic data. For these species, the statistical error starts to be significant once the abundance, which typically decreases, is modified by the chemistry. This is the case for CH$_4$, H$_2$CO, CH$_3$OH, H$_2$S, OCS and S. Second or later generations of compounds are affected by the kinetic data uncertainties from the beginning, and the errors for their abundances generally increase with increasing generational number. This effect can be explained by the increasing number of reactions and species involved in their formation. The increase of the error with the generational rank can, however, be verified only for species storing a small fraction of the global reservoir of the chemical elements of which they consist. Indeed, because the elemental abundances in the gas phase are fixed, the sum of the abundances of all species bearing a given element represents a constant reservoir that is not affected by uncertainties due to kinetic data. For instance, $\Delta$[Sulphur]=0 , where the elemental abundance of sulphur is given by
\begin{equation}
\rm [Sulphur] = [S] + [SO] + [SO_2] + [CS] + [HS] + [H_2S] + \ldots .
\end{equation}
For $t<10^4$~yr, atomic S is the dominant S-bearing species by two orders of magnitude, and the error of its abundance is thus negligible. At $t>10^5$~yr, most of the sulphur in the gas phase is in the form of SO$_2$ and, despite being a late generation molecule, the error on its abundance becomes insignificant.
In their study of error propagation in the chemical modeling of interstellar molecular clouds, \citet{2004AstL...30..566V} found a correlation between the error and the 
number of atoms per species. We did not find such correlation in the hot corino chemistry. This might be due to differences in the physical conditions but also to the fact that  we used a reduced sample of species. Ninety percent of the species included in our model have four atoms or fewer, and only four molecules possess six or seven atoms. Therefore, we cannot determine if such a correlation exists for the complex species with seven or more atoms, although it seems likely since these species are expected to have low abundances and be of late generational rank.  

For the same physical conditions, the average molecular abundance, $\overline{X}$, is normally close to the standard abundance $X_0$ (computed with the standard rates). However, for some species, a factor of up to 2 difference is found between the two values, especially when the distribution of $X$ diverges from a log-normal distribution. This is for instance the case for H$_2$S (as seen on Fig.~\ref{ab_mean_ref}) around t=$6\times 10^4$~yr.  On the same figure, we can see that the fast decrease of the H$_2$S  abundance produces sharp spikes in the ratio $\overline{X}/X_0$, which occur when the curves of $\overline{X}$ and $X_0$ cross and re-cross each other.

\subsection{Uncertainties in abundance ratios}

\begin{figure}
\centering
\includegraphics[width=1\linewidth]{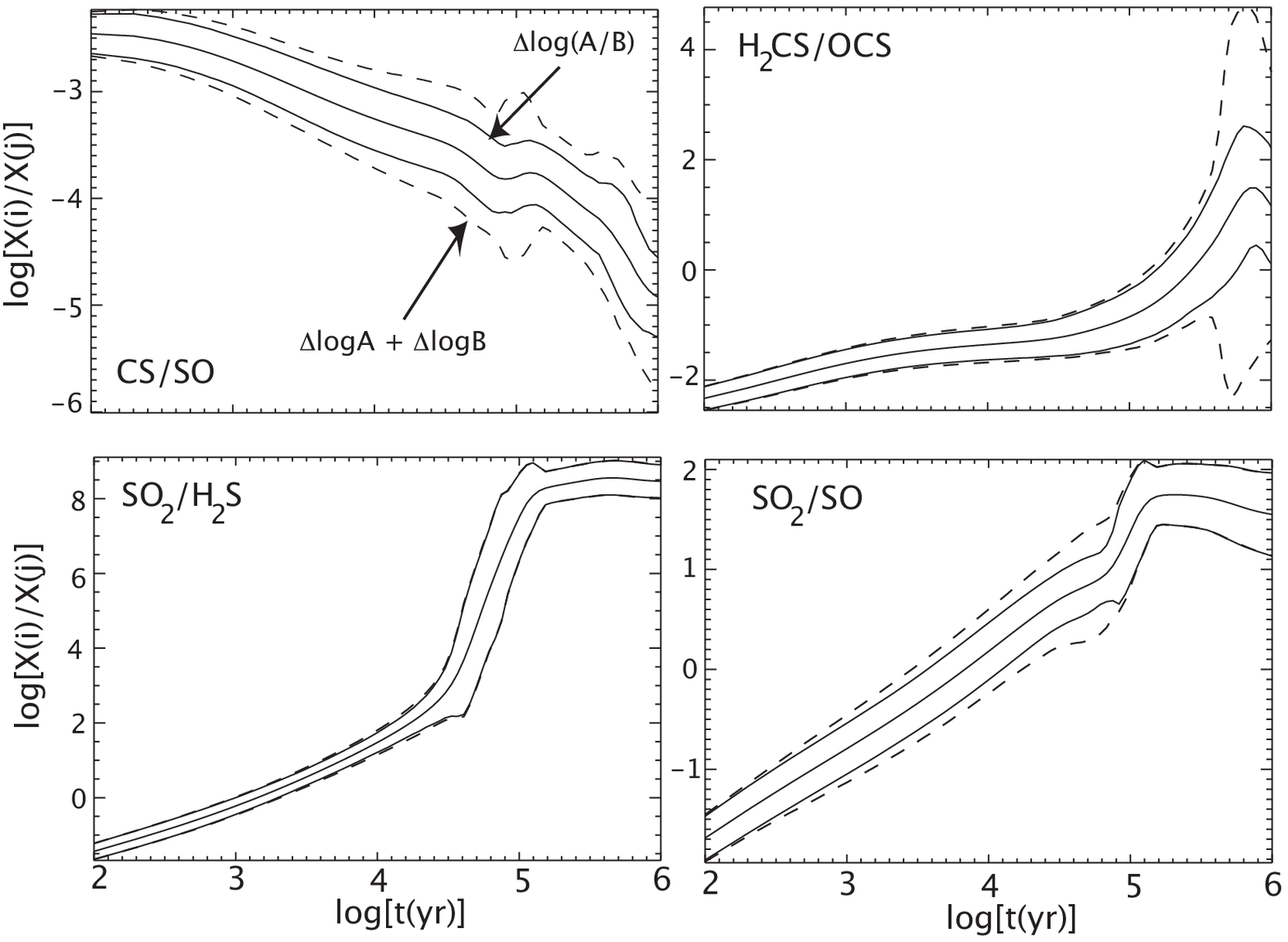}
\caption{Abundance ratios (solid central lines) with error bars computed using two methods: $\Delta \log(A/B)$ (solid lines) and $\Delta \log A + \Delta \log B$ (dashed lines).}
\label{error_rap}
\end{figure}

The uncertainty in abundance ratios can be computed using two different approaches. The first one involves the calculation of the abundance ratios for each run and the determination of the error $\Delta \log(A/B)$. The second method involves the error for each abundance and their subsequent addition: $\Delta[\log A] + \Delta[\log B]$. 
Fig.~\ref{error_rap} shows the evolution of the logarithm of some interesting abundance ratios with errors computed using both methods. In the absence of correlation between A and B, the two resulting errors are similar. This is the case for SO$_2$ and H$_2$S because, in our model,  SO$_2$ is formed by the oxygenation of the initial atomic sulphur more than from the destruction of H$_2$S. Another example is the ratio H$_2$CS/OCS before $10^4$~yr since OCS only starts to be destroyed after this time to form H$_2$CS. When the species A and B are chemically related, $\Delta[\log A] + \Delta[\log B]$ overestimates the error on the ratio. The abundance ratios CS/SO and SO$_2$/SO are two good examples since the formation mechanism of CS and SO$_2$ are directly related to SO. For H$_2$CS/OCS the discrepancy becomes huge for integration times between $10^5$ and $10^6$~yr: $\Delta[\log A] + \Delta[\log B]$ covers 7 orders of magnitude while $\Delta[\log(A/B)]$ remains between 2 orders of magnitude. As a consequence in this paper, we use the error computed for the abundance ratios $\Delta[\log(A/B)]$.

\section{Some consequences for hot core models} 

\subsection{The cosmic-ray ionization rate in IRAS16293-2422}\label{Res3}

\begin{figure*}
\begin{center}
\includegraphics[width=0.8\linewidth]{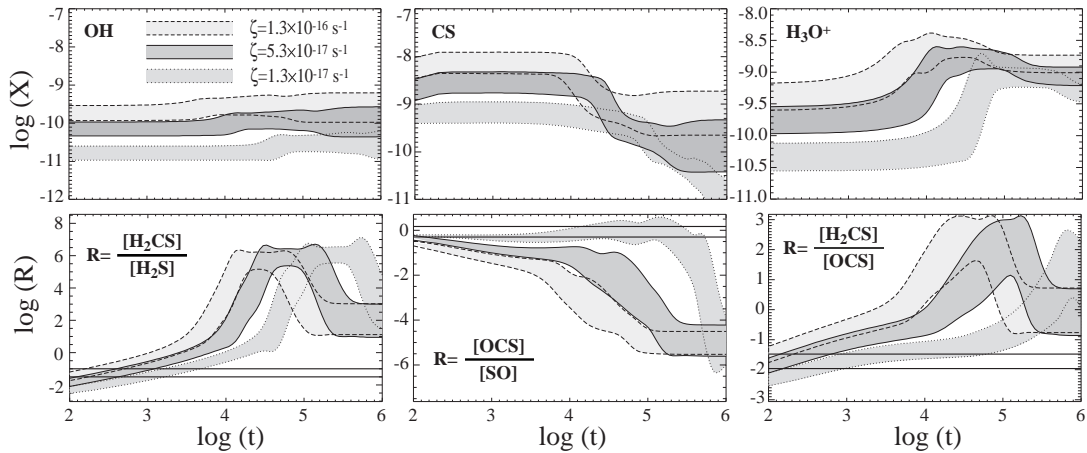}
\caption{The upper panel presents calculated abundances (with respect to H$_2$) of OH, CS and H$_3$O$^+$ as a function of time with an error bar (95.4\% confidence level) for three different ionization rates. The lower panel presents three calculated abundance ratios (H$_2$CS/H$_2$S, OCS/SO and H$_2$CS/OCS) with similarly estimated uncertainties for the three cases. The solid lines in the lower panel represent the ratios observed in IRAS16293-2422 taking into account the uncertainties in the observations.}
\label{6fig_3Z}
\end{center}
\end{figure*}

The ionization rate $\zeta$ is an important parameter of chemical models but is weakly constrained. It depends on the cosmic-ray flux, which cannot be directly determined. Some constraints have been derived from the comparison of observed abundances with numerical predictions done with several values of $\zeta$ \citep{1979ApJ...234..876W,1998ApJ...499..234C,2000A&A...358L..79V,2002ApJ...565..344C,2004A&A...418.1021D}. None of the models used included any uncertainties in the rate coefficients. Here we address the question whether or not it is possible to distinguish among several ionization rates in hot cores.

For that purpose, we ran models for hot cores (100~K and $10^7$~cm$^{-3}$) with three different values of the ionization rate: $\zeta = 1.3\times 10^{-17}$, $5.3\times 10^{-17}$ and $1.3\times 10^{-16}$~s$^{-1}$. Then, we searched for those species with abundance or abundance ratio that would allow us to distinguish among the values of $\zeta$ despite the dispersion due to kinetic uncertainties. 
Our results show that the best molecules to constrain the ionization rate in hot cores appear to be OH, CS and H$_3$O$^{+}$ (see Fig.~\ref{6fig_3Z}). Indeed, the abundance of OH remains almost constant and the abundances of CS and H$_3$O$^{+}$ do not vary much with time before $10^4$~yr. While OH can be detected at a large number of frequencies under 100~GHz thanks to its lambda-doubling transitions, H$_3$O$^+$ can be detected via rotational-inversion transitions at significantly higher frequencies. 
If the age of the source is known, the OCS, H$_2$S, and SO molecules can also be used.  On the contrary, the species H$_2$CS, CH$_3$OH, H$_2$CO, O$_2$ and SO$_2$ show distinct but close abundance distributions only for the most extreme rates $1.3\times 10^{-17}$ and $1.3\times 10^{-16}$~s$^{-1}$ and intermediate values of the rate can not be distinguished. Other species, like H$_2$O, CH$_4$, C$_2$H$_2$, CO and O are less sensitive to $\zeta$ at young ages ($t<10^{5}$~yrs) and exhibit strong overlaps after.

From both the theoretical and observational points of view, the use of abundance ratios instead of individual abundances relative to H$_2$ involves less uncertainty. In particular, observed abundance ratios avoid the uncertainties on the H$_2$ column density  
and the size of the source, assuming that both molecules of the ratio come from the same region. Also, theoretical abundance ratios are less sensitive to the initial conditions, which are usually not well constrained. Our results show that some of the commonly used abundance ratios involving sulphur-bearing species can then be used without confusion to distinguish among the three cosmic-ray ionization rates. As an example, we show the evolution of three of the abundance ratios: H$_2$CS/H$_2$S, OCS/SO and H$_2$CS/OCS for the three values of $\zeta$ in Fig.~\ref{6fig_3Z}. 

The abundance of OH has never been determined in any hot core, to the best of our knowledge.  Moreover, there exist only observations of low-energy transitions of CS ($E_{up}  < 45$~cm$^{-1}$), which do not allow determination of its abundance in the inner regions of protostellar objects \citep{2002A&A...390.1001S}. CS transitions with $J>7$ cannot be observed with current ground based telescopes because of the receivers frequency ranges or they are absorbed by the atmosphere. There was one attempt to detect H$_3$O$^+$ towards IRAS16293-2422 by \citet{1992ApJ...399..533P} with no success. The authors deduced an upper limit of $2\times 10^{-10}$ for the abundance of this molecule which would strongly constrain the ionization rate in this source. However, this limit is for a $18''$ beam size. If we consider a 2$''$ source size for the hot core of IRAS16293-2422 and a H$_2$ column density of $7.5\times 10^{22}$~cm$^{-2}$ \citep[see][]{2000A&A...357L...9C}, the new limit on the H$_3$O$^+$ abundance (compared with H$_2$), assuming LTE, is $4\times 10^{-8}$. The limit is thus too high to conclude anything when comparing with Fig.~\ref{6fig_3Z}.  Note that in hot cores, the typical "ionization tracers" \citep[i.e. HCO$^+$ and CCH, ][]{1997ApJ...481..296Y,1997A&A...318..579G} are not abundant. Our attention thus shifts to the ratios of the sulphur-bearing species discussed above.

In Fig.~\ref{6fig_3Z}, we compare computed values of the ratios and their error bars with observed ratios and uncertainties for the low mass protostellar source IRAS16293-2422 \citep{2002A&A...390.1001S,2004A&A...413..609W}. The observed OCS/SO ratio clearly constrains the cosmic-ray ionization rate to be $1.3\times 10^{-17}$~s$^{-1}$. The two other observed ratios H$_2$CS/H$_2$S and H$_2$CS/OCS are in agreement with this rate for ages around $10^3$~yr. As already noticed by \citet{2004A&A...422..159W}, this result is contradictory to \citet{2004A&A...418.1021D}, who could reproduce the abundances towards this source only with a high cosmic-ray ionization rate of $1.3\times 10^{-16}$~s$^{-1}$ \citep[see][for discussion]{2004A&A...422..159W}. 
Note that both the rate and the age were already determined by \citet{2004A&A...422..159W} and the goal here is to show that even with the introduction of the uncertainties in the reaction rates, we can still distinguish  ionization rates between $1.3\times 10^{-17}$ and $1.3\times 10^{-16}$~s$^{-1}$. In other words,  molecular abundances can be used to constrain the ionization rate, but one needs to verify that  uncertainties in the reaction rate coefficients do not confuse the results. 

\subsection{Consequence on the age of IRAS16293-2422 hot corino}

\begin{figure*}
\centering
\includegraphics[width=0.9\linewidth]{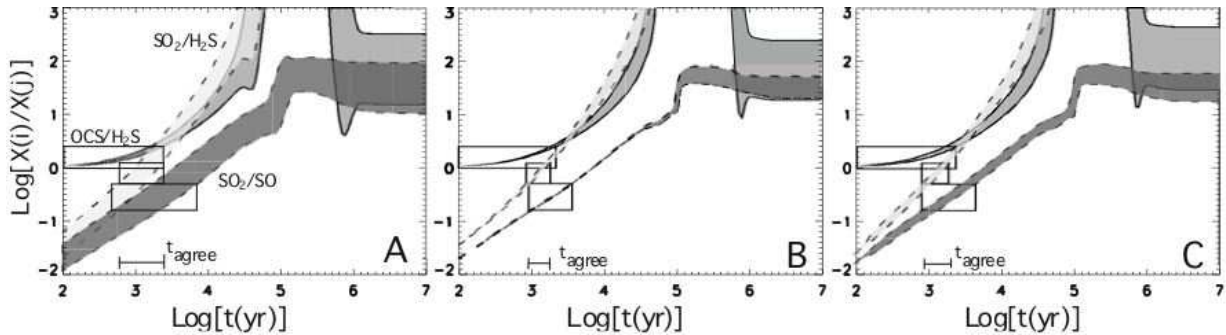}
\caption{Comparison between the observed abundance ratios SO$_2$/SO, SO$_2$/H$_2$S and OCS/H$_2$S (black empty boxes) towards IRAS16293-2422 and the theoretical predictions (gray contours). A: including the actual uncertainties in rate coefficients; B: decreasing the uncertainties of the 5 reactions in Table~\ref{reac_imp} to 0; C: decreasing these uncertainties  to achievable values (see Table~\ref{reac_imp}). }
\label{R0_R5}
\end{figure*}

\begin{table*}
\caption{List of important reactions.\label{reac_imp}}
\begin{tabular}{llllll}
\hline
Reaction & $\alpha$ & $\beta$ & $\gamma$ & UMIST uncertainties & Achievable uncertainties\\
\hline
\hline
O + SO $\rightarrow$ SO$_2$ + h$\nu$ & 3.20e-16 &  -1.50   &  0.0 & Error within factor 2 & 20\% \\
H$_2$ + CR$^*$ $\rightarrow$ H$_2^+$ + e$^-$ & 1.30e-17 &   0.00    &   0.0 & Error within
factor 2$^{***}$ & 20\% \\
H$_2$O + CRPHOT$^{**}$ $\rightarrow$ OH + H & 1.30e-17 &   0.00  &   971.0 & Error within factor 2 & 20\% \\
CO + S $\rightarrow$ OCS + h$\nu$ & 1.60e-17 &  -1.50  &     0.0 & Error $< 25$ \% & 25\% \\
H$_3^+$ + S $\rightarrow$ HS$^+$ + H$_2$ & 2.60e-09  &  0.00   &    0.0 & Error within
factor 2 &10\% \\
\hline
\multicolumn{6}{l}{$^{*}$Cosmic-ray ionization; $^{**}$Cosmic-ray-induced photodissociation}\\
\multicolumn{6}{l}{$^{***}$Uncertainty based on laboratory experiments rather than uncertainty in the cosmic-ray ionization rate}\\ 
\end{tabular}
\end{table*}

\citet{2004A&A...422..159W} constrained the age of the IRAS16293-2422 hot corino by comparing the SO$_2$/SO, SO$_2$/H$_2$S and OCS/H$_2$S abundance ratios observed towards the source with the predictions of their chemical model. In their comparison, they only took into account the errors in the observed ratios, which are due to the calibration error of the telecopes. \citet{2004A&A...422..159W} obtained an age of $(2\pm 0.5)\times 10^3$~yr, where the
uncertainty in the age, 25\% \citep[see Fig.~7 of][]{2004A&A...422..159W}. We did the same comparison including the uncertainty in the rate coefficients with the method described in \S~\ref{method}.
 In Fig.~\ref{R0_R5}, panel A, we report the new comparison including both the uncertainty in the observations and in the modeling. In this case, we obtain an age between 800 and $2,500$~yr, $(1.65\pm 0.85)\times 10^3$~yr, implying an error of 50\% in the age\footnote{The age and its relevant error are defined as follows: let us define  $\Delta t= [t_{min},t_{max}]$ as the interval of time where the three observed ratios are reproduced within the uncertainties by the model ($\Delta t = t_{agree}$ in Fig.~\ref{R0_R5}), then the age is $t_{min} + \frac{\Delta t}{2}$ and the error is $\frac{\Delta t}{2}$. }. As already noticed in \citet{2004A&A...422..159W}, this age is relatively short compared with previous estimates \citep[$\sim 10^4$~yr][]{2000A&A...357L...9C,2002A&A...395..573M,2004A&A...413..609W}. However, this determined age represents only the time from the evaporation of the mantles; the dynamical time needed to reach the required luminosity to form the observed hot core is an additional $\sim 3\times 10^4$~yr.

Using the sensitivity analysis described in \S~\ref{method}, we then determined the reactions that produce most of the uncertainties in the three ratios (SO$_2$/SO, SO$_2$/H$_2$S and OCS/H$_2$S) at an age of $10^3-10^4$~yr. We found five important  reactions,  which are summarized in Table~\ref{reac_imp} with their rate coefficients ($\alpha$. $\beta$, $\gamma$ from the UMIST database) and their uncertainties.  The reactions are given in decreasing order of importance. Note that reactions 2 and 3 are directly related to the cosmic-ray ionization rate studied in \S~\ref{Res3}, and their uncertainties are based at least partially on astronomical rather than laboratory considerations. 
To check the importance of these five reactions, we decreased their uncertainties to 0 and to achievable values (see  Table~\ref{reac_imp}) as shown on Fig.~\ref{R0_R5} (panels B and C). In estimating achievable values, we were helped by Nigel Adams (private communication).  The uncertainties for the radiative association reactions between neutral species are rather speculative and contain the assumption that experiments to determine the rate coefficients can have the same low uncertainty as other experiments in neutral-neutral reactions. The uncertainty in panels B and C of Fig.~\ref{R0_R5} is still high at later times ($\sim 10^6$~yr) because other reactions than the ones listed in Table~\ref{reac_imp} are important. 
%ERIC: PLEASE COULD YOU ADD SOME FEW SENTENCES ON THE CHOSEN ACHIEVABLE %UNCERTAINTIES?
%Decreasing the error on these rates to 0 decreases the relative errors on the ratios from 0.16, 0.14 and 0.09 to 0.006, 0.08 and 0.06 for SO$_2$/SO, SO$_2$/H$_2$S and OCS/H$_2$S respectively (at t=$10^4$~yr). After $10^5$~yr, the relative errors increase again especially for OCS/H$_2$S because other reactions influence the ratios. Now, if we consider achievable uncertainties as given in Table~\ref{reac_imp}, the relative error on the abundance ratios can be as low as 0.05, 0.08 and 0.07 (for  SO$_2$/SO, SO$_2$/H$_2$S and OCS/H$_2$S respectively). In the first case, we obtain an age of $(1.33 \pm 0.44) \times 10^3$~yr so an uncertainty of 33\% whereas an achievable error on the rates give an age of $(1.44 \pm 0.55) \times 10^3$~yr (38\% of uncertainty).
Table~\ref{Res_error} gives the range of the computed abundance ratios SO$_2$/SO, SO$_2$/H$_2$S, OCS/H$_2$S at $10^4$~yr using the current (col.~1), null (col.~2) and achievable (col.~3) uncertainties for the five reactions listed in Table~\ref{reac_imp}. Note that in the case of these abundance ratios, we found that the relative error is well determined by the $2\sigma$ dispersion (see \S~\ref{def_error}). In the last line of the table, we report the range of age found for the IRAS16293-2422 hot core. 

\begin{table}
\centering
\caption{Range of computed abundance ratios and age of IRAS16293-2422 hot corino at t=$10^4$~yr. \label{Res_error}}
\begin{tabular}{lccc}
\hline
 & \multicolumn{3}{c}{[minimum value, maximum value]} \\
 & current & null & achievable \\ 
\hline
\hline
SO$_2$/SO & [1.12, 2.34]  &  [1.60, 1.64]  &  [1.45, 1.82] \\
SO$_2$/H$_2$S & [25.1, 47.9]  &  [28.8, 41.7]  &  [28.8, 41.7] \\
OCS/H$_2$S & [11.2, 17.0]  &  [12.0, 15.8]  &  [11.7, 16.2] \\
 Age (yr) & [800, 2500] & [890, 1770] & [890, 1990]  \\
\hline
\end{tabular}
\end{table}

\section{Conclusions}

We report in this article a study of the impact of rate coefficient uncertainties on pseudo-time dependent models and some methods to improve the accuracy of theoretical predictions. We first present a general method to include the treatment of rate imprecisions in models and a procedure to identify the main uncertain reactions responsible for the error on the abundances. 
We found that for hot core models the relative errors $\frac{\Delta X\pm}{X}$ on the predicted abundances of the more abundant species are lower than 50\% (with a 95.4\% level of confidence) for times earlier than $10^4$~yr, which is comparable to the uncertainties in observed abundances. This error depends on time, H$_2$ density, and temperature, and is related to the generational rank of the molecule in the hot core chemistry. The first generation of molecules, which are evaporated from the grains, has lower uncertainties than the second and later generations. 

We also studied the consequences of rate uncertainties on the use of models to constrain the ionization rate $\zeta$ and the age of hot cores. Among all the molecules, we found that OH, CS and H$_3$O$^+$ were the best to distinguish among $\zeta = 1.3\times 10^{-17}$, $5.3\times 10^{-17}$ and $1.3\times 10^{-16}$~s$^{-1}$. Indeed, these molecules have reasonable abundances ($>10^{-11}$) which do not vary much with time before $10^4$~yr and depend on $\zeta$. Sulphur bearing abundance ratios such as H$_2$CS/H$_2$S, OCS/SO and H$_2$CS/OCS can also be used since these molecules are easily observed in hot cores and show strong dependence on $\zeta$. 

We compared our modeling including the uncertainties with observed S-bearing abundance ratios in order to constrain the age of IRAS16293-2422, as \citet{2004A&A...422..159W} did without the rate uncertainties. We also developed a simple method to sort the reactions by the uncertainty they produce in the model results. This allowed us to identify a group of 5 reactions mainly responsible for the error in the SO$_2$/SO, OCS/H$_2$S and SO$_2$/H$_2$S ratios, at integration times between $10^3$ and $10^4$~yr. The decrease of the actual error for these 5 reactions to achievable values would decrease the uncertainty of the age of this hot core from $\pm$50\% to $\pm$38\%.

In conclusion, for young hot cores ($\le 10^4$~yr), the modeling uncertainties related to rate coefficients are reasonable and comparisons with observations make sense. On the contrary, the modeling of older hot cores is characterized by strong uncertainties for some of the important species. In both cases, it is crucial to take into account these imprecisions to draw conclusions from the comparison of observations with chemical models. In addition, being able to identify, among the thousands of reactions  involved in interstellar chemical networks, the few reactions for which a high accuracy is required can be useful especially for old regions. Studies such ours rely on the estimation of the uncertainties provided by kinetic databases and thus, an effort should be done to better quantify and the rate coefficient uncertainties.

\begin{acknowledgements}
V. W. and F.S. wish to thank Michel Dobrijevic for fruitful discussions on error propagation in chemical networks.
V. W. and E. H. thank the National Science Foundation for its partial support of this work. 

\end{acknowledgements}

\end{document}